# A simple and straightforward technique for single shot measurement of terahertz profile

Sonal Saxena*[1,2], B.S.Rao[1], J.A.Chakera[1,2], P.A.Naik[1,2]
1 Laser Plasma Division, Raja Ramanna Centre for Advanced Technology, Indore 452 013
2 Homi Bhabha National Institute, Mumbai 400092
*Email: sonals@rrcat.gov.in

**Abstract:**

A new detection scheme capable of acquiring the entire spatiotemporal profile of terahertz radiation in a single laser shot is being demonstrated. The design allows temporal resolution of the order of transform-limited pulse duration of the probe, which is an absolute benefit over the compromised resolution of otherwise prevalent single-shot detection schemes based on spectral encoding of terahertz waveform on a temporally chirped readout pulse. This makes the technique perfectly suitable for sensitive spectroscopy studies. The single shot detection technique presented here relies on space-to-time mapping of terahertz temporal profile by use of a converging probe intensity front. The present approach does not require any specialized optics and is implemented using very straightforward alignment procedure. It has shown to reproduce the temporal waveform of terahertz radiation faithfully.

**Introduction:**

The region between far infrared and microwaves in the electromagnetic wave spectrum possesses some unique properties and is termed as the terahertz (THz) gap, as it was inaccessible either by the photonic or electronic mechanisms [1, 2]. The last few decades saw the advent of bright THz sources and sensitive detectors, which allowed the realization of already proposed and newly perceived applications in this frequency range [3]. One of the major applications involves the study of several unexplored phenomena and material properties, in this part of spectrum using the THz time-domain spectroscopy (THz-TDS) [4]. Full time dependent waveform of the THz electric field is analysed for spectral features and was possible only because of the evolution of ultra-short duration pulsed lasers. This was an important accomplishment as THz detection was limited to thermal absorption based devices giving integrated or average power of the radiation only. Complete characterization of the THz radiation, including amplitude and phase information is executed by scanning a few femtosecond duration, readout pulse across the THz pulse of the order of picoseconds in time to reproduce the complete temporal evolution of THz electric field. Scanning techniques generally employ photoconductive antenna [5] or the properties of electro-optic crystals [6]. Electro-optic sampling (EOS) is a much more commonly used technique because of a flat frequency response over a wider range, in comparison to the carrier lifetime and design dependent spectral response of photoconductive antenna in addition to several other beneficial features [7].

Many of the THz sources are based on ultrashort pulse duration lasers and also high energy lasers are used to generate gas plasma to act as large bandwidth terahertz source and detector. These plasmas are seen as an attractive and promising THz source as larger laser energies can be coupled into the medium in comparison to damage threshold and bandwidth issues related to material media [8]. Scanning detection schemes are based on probing in time, with a part of the same laser used for generation across the THz pulse duration and is based on the assumption that laser energy delivered in each shot is the same . This coherent detection mechanism will not be feasible with high power lasers, which suffer from large shot to shot fluctuations. These ultrahigh intensity lasers come with low repetition rate, which implies even longer data acquisition time. In addition to the generation and detection, applications too have grown more demanding in terms of accuracy and time taken to record a single spectral axis. Earlier pump-probe geometries studied the linear terahertz response of a system at equilibrium and now are capable of reading dynamics of reversible processes following photo-excitation, like carrier dynamics in semiconductors by employing more than one temporal delay line [9]. If recording a single temporal axis takes n data points for effectively extracting the spectrum from waveform, then a two-dimensional plot will require $n^2$ laser shots. Experiments are also done to deduce the extent of coupling between THz-resonant degrees of freedom in a system. Vibrational, electronic or rotational couplings, in the THz range can be deduced through a study of pump-induced response of a system measured as the time-evolution of the change in THz probe spectrum. This technique known as Multidimensional THz-TDS is an evolving one [10] and has been used to study the coupling between electronic degrees of freedom in quantum wells [11]. These new studies are very exciting, but shall scale up the time taken for measurement by orders of magnitude and practical implementation will be feasible only if one temporal axis is registered in one laser shot and then averaged for improving accuracy. While reading reversible phenomena demands waveform detection on a faster time scale, irreversible phenomena make the demand for a scan-free, accurate single-shot waveform acquisition a prerequisite. The desire to record rapid time dynamics of irreversible phenomena like material damage, phase transitions or chemical changes has driven the growth of single-shot spectrum detection methods in the THz frequency range.

A few single shot THz waveform detection methods have already been proposed, most of which rely on the spectral encoding of THz profile on a sufficiently long temporally chirped probe pulse [12]. The basic mechanism involved is the electro-optic technique only, where the THz field amplitude is imprinted as the probe pulse's polarization change. The readout pulse is made to pass through an analyser to record the change in intensity of one of the either polarization states. In spectral-encoding methods, an intensity modulated spectrum is recorded on the spectrograph and frequency-to-time calibration of the reference probe pulse is utilized to extract the temporal structure of THz electric field. An additional system for introduction of temporal chirp in the laser beam is required and the knowledge of its exact frequency-to-time calibration is crucial. Though calibrated stretchers are regularly used in laser laboratories, this method of detection through spectral encoding suffers from a fundamental limit on the achievable time resolution. While the time window increases for larger chirp, time resolution [13] given by $\sqrt{\tau_0 \cdot \tau_{ch}}$, where $\tau_0$ is transform-limited pulse duration and $\tau_{ch}$ is the chirped pulse duration, gets degraded. High distortion accompanies the signal when time resolution is insufficient for monitoring fast processes [14]. Optimization of the resolution and chirp rate has to be done for different THz pulse durations. An improvement in this method is done by using the chirped readout pulse along with a short transform-limited duration readout pulse for linear interferometry [15]. Both the pulses are matched spatially on the electro-optic crystal with a relative time delay of δt and sent to a spectrograph. The THz signal is encoded on the chirped pulse and the spectral interference, when compared with a reference interferogram without the signal can be used to extract the temporal waveform of THz. The time resolution is given by the $\tau_0$ and the temporal window is decided by $\tau_{ch}$ and the spectral resolution of spectrometer. In addition to the requirement of a high-resolution spectrometer, very precise alignment is needed for larger temporal windows. Since interferometry is a highly sensitive technique, stringent experimental conditions are critical for maintaining the fringe fidelity. The aforementioned techniques can broadly be classified as frequency-to-time mapping schemes. For the space-time encoding schemes a tilted optical readout intensity front is made to interact with an enlarged THz spot size at the electro-optic crystal for attaining larger time windows. The probe samples the THz field in subsequent sections of time, as they coincide spatially on the detector crystal. The intensity front of probe beam is inclined using a prism or grating [16, 17], or non-collinear crossing [18]. Though these schemes have been appreciated for technical simplicity and easy alignment, the reduction in THz intensity is a big drawback. The decrease in THz field intensity reduces the signal with respect to the background level and hence the method is not suitable for sensitive spectroscopy studies. Another approach based on custom made dual transmission echelons works on the principle of angle-to-time mapping [19]. Though the method involves minimal modification in probe path, thickness and quality of fabricated optics along with the focussing of beam-let array pose restrictions on implementation of the technique.

In the following sections, a new single shot detection technique is discussed and its merits and demerits are compared as those against the established ones. The methods will be analysed in terms of time resolution, temporal window scanned in one shot, ease of implementation, accurate replication of THz electric field and the possibility of implementing balanced detection. For the purpose of demonstration, two colour laser generated air plasma has been used as the THz source. This detection scheme is equally suitable for detecting THz radiation from any source.

**Experimental Setup:**

The experiment was conducted using 10 TW Ti: sapphire laser system at Laser Plasma Division, Raja Ramanna Centre for Advanced Technology, Indore. The system delivers laser pulses at 800 nm central wavelength and 20 nm bandwidth with a minimum of 45 femtosecond duration at 10 Hz repetition rate. The laser radiation was divided into two parts. The larger portion was utilized for THz generation based on the photo-ionization of air using fundamental and second harmonic radiation with some phase difference between the two, i.e. the two colour laser air-plasma source [20]. A finite phase difference between $\omega$ and $2\omega$ results in an asymmetric electric field in the plasma. This field causes the electrons to get drifted in one direction more than the other, leading to a net transient current on the time scale of pump-pulses' envelope which drives the THz fields. The p-polarized laser was focused using a 70 cm focal length lens and a 100 µm thick β-Barium Borate (BBO) was placed before the focus for second harmonic generation. A very small part of the main laser beam, reflected from a glass wedge was used as the probe pulse. A 90° off-axis parabola mirror (OAPM) was placed at a distance equal to its focal length form the plasma filament to collimate the conical THz emission. The probe beam was passed through a delay stage to the surface of HRFZ-Si (High Resistance Float Zone Silicon) wafer, used as a broadband THz filter and for combining the paths of THz and probe. Collimated THz beam and the probe beam were focussed using the same optics, i.e. the second OAPM, on the surface of a 200 µm thick ZnTe crystal cut in <110> plane. The electro-optic properties of ZnTe are exploited here for detection. Initially, the THz waveform was traced using the conventional mechanical delay line and then compared with the results obtained from single-shot method.

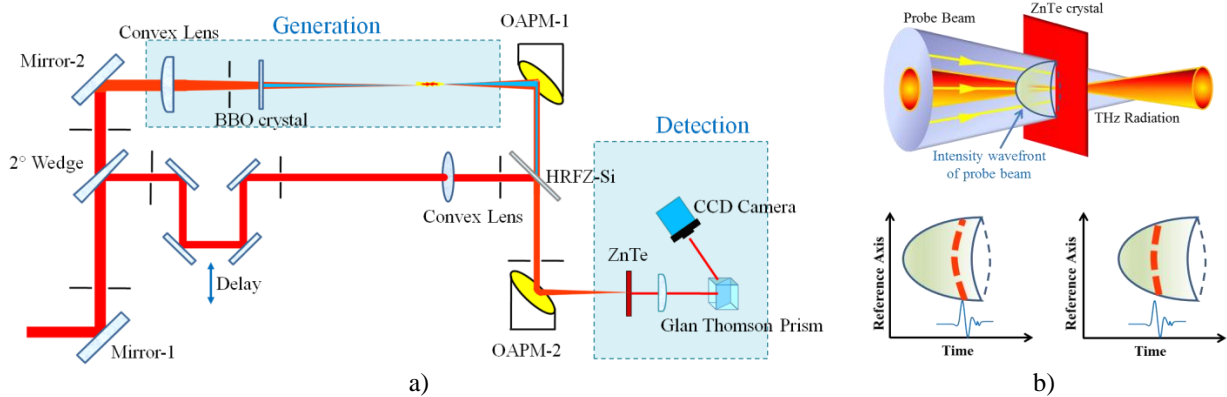

*Fig.1: a) The schematic experimental setup showing generation and detection mechanism. b) The change in delay of probe beam leads to the THz peak field getting imprinted on a different part of the probe each time.*

A schematic of the experimental setup for generation and detection of THz radiation is shown in fig. 1.a). For the implementation of single shot detection scheme, a small focal length convex lens was placed in the probe path, before the HRFZ-Si wafer. The purpose of this lens is to shift the focus of probe beam after the second OAPM to a point farther than the surface of ZnTe crystal. This results in a larger spatial spot size of laser beam with respect to the THz focal spot. A converging laser intensity front falls on the detector crystal and experiences the THz field intensity in an extended stretch of time. Hence the complete temporal information of THz electric field is recorded as a spatially and temporally integrated image of the ZnTe crystal surface on a ccd (charge coupled device) camera. The THz radiation is generated at a fixed time with respect to the laser shot. As the arrival of THz pulse at the detector crystal is constant in time, changing the delay in probe beam will vary the region of their interaction each time. The reason for variation in the recorded image due to change of probe delay is depicted pictorially in fig.1.b). The change in position of THz peak intensity in the pixel matrix of ccd detector, with respect to the probe delay serves as the pixel/distance-to-time calibration for extracting the temporal waveform of the radiation. Optical geometry of the probe laser decides the time window scanned in one laser shot, which is 4 ps in the present setup.

**Results and discussion:**

The converging intensity front of probe beam interacts with the spatially focussed spot of the THz radiation in subsequent time sections. The marginal rays interact with the THz field prior to those towards the axial direction. This way images radially symmetric about the centre of probe spot are captured on the camera. We imaged the region of interaction from the rejected side of the Glan-Thomson prism. With the use of a quarter wave-plate in the probe path to circularly polarize the incident beam, images from both sides could be captured and subtracted to achieve better signal-to-noise ratio. Images at different probe delays were recorded to get the space-to-time calibration. The background intensity due to probe beam was subtracted in each image by recording images without THz radiation at every time delay. '0' femtosecond is an arbitrary reference point in time and delay was introduced with respect to that point. The least count of the delay stage used in our experiment was 20 μm or 66 fs, so images after 3, 6 and 9 rotations are shown in fig.2.

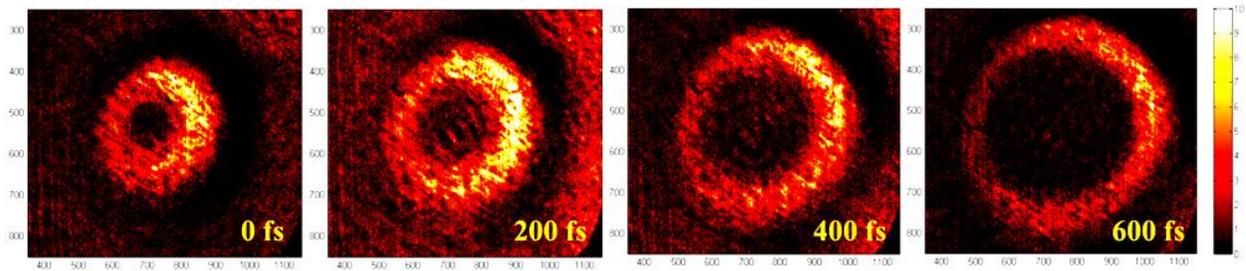

*Fig. 2: Images of terahertz spatial profile taken at different probe delays*

As the temporal information of THz radiation is inscribed on the probe beam in a radially symmetric manner, any line cut passing through the centre of probe beam's spatially integrated image will be the same. We have selected a vertical line cut, as the asymmetry in horizontal direction seen in the image is because of the source characteristics [21].The pixel number corresponding to the THz peak intensity in a line cut on one of the either side is registered and its shift from the centre with changing time delay in probe is plotted. The trace is given fig. 3.a). Though the probe beam's contour is effectively measured using the optical geometry of probe path, this plot gives the calibration with better accuracy. Now, a particular probe beam delay was fixed such that all the oscillations before and after the THz peak are effectively recorded and all the data is taken in this configuration. Applying the pixel-to-time calibration to a line cut through centre of the image gives the field v/s time substructure for the THz radiation. As probe beam got reflected from the surfaces of silicon wafer and OAPM and got transmitted through the ZnTe crystal, its intensity front suffered microscopic distortions riding on the macroscopic shape. Spatial inhomogeneity in the probe beam was corrected by subtracting the image taken in absence of THz field.

A comparison of the waveforms obtained through conventional EOS and the converging probe, single-shot detection techniques are shown in fig. 3.b). In our case the time resolution of EOS is given by the least count of delay stage, i.e. 66fs and the single-shot method is the transform-limited probe pulse duration, i.e. 45 fs.

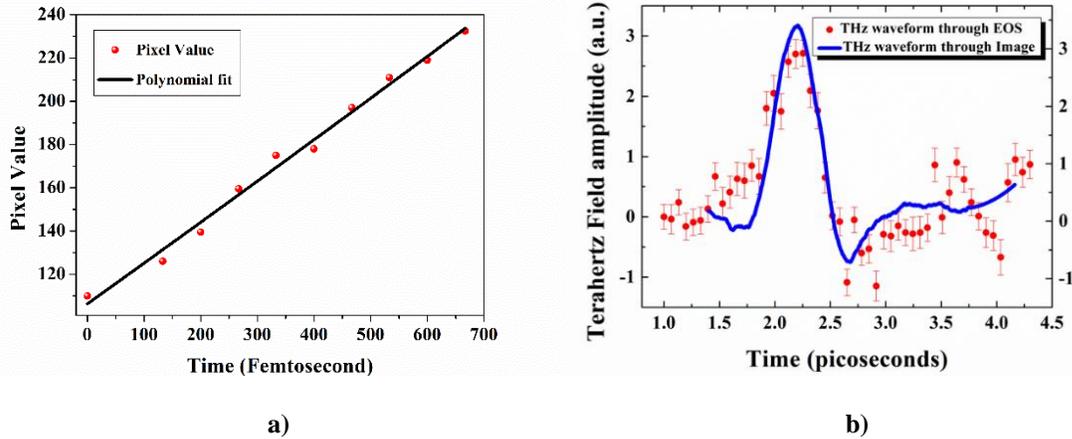

a)            b)

*Fig. 3: a) Plot of the pixel value corresponding to the THz peak intensity in one arm of line cuts through images at different probe delays. b) The terahertz profiles obtained from the conventional EOS method and the single-shot method using curved probe intensity front match well.*

As inferred from the above discussions, the single-detection method using a curved intensity front is much simpler to implement than the alternate methods. In contrast to the other space-to-time mapping techniques, balanced detection can be easily applied and reading a focussed THz spot promises a better signal-to-noise ratio. The time resolution is given by the transform-limited duration of the probe pulse and is affected by factors common to any electro-optic technique. The presented design allows for easy switching between single-shot and scanning detection for verification purpose, without effecting the generation. A major benefit of this method is that the magnitude of time window can be regulated by the geometry of optics used. The calibration mentioned above for deciphering the shape of probe intensity front is not a very involved procedure and has to performed only once for a complete set of data taken without disturbing the probe path geometry.

**Conclusion:**

A new THz profile detection method based on space-to-time grading has been proposed and evaluated with respect to the other methods devised and used. Merits of the scheme based on several parameters were also judged and found that without the requirement of any specially designed optics or precise alignment the method reproduces the THz profile faithfully. The scheme matches well with the conventional EOS on all grounds, while reducing the time taken for data acquisition by a huge amount. Spatial profile of the THz radiation is imaged and shot-to-shot fluctuations can be suppressed implementing the balanced detection. Overall this technique offers a decent option for real time data acquisition for THz radiation and enables several applications efficiently.